\documentclass[nonatbib,preprint,nocopyrightspace]{sigplanconf}
\usepackage{amsmath}
\usepackage{amstext}
\usepackage{amssymb}
\usepackage{amsthm}
\usepackage[T1]{fontenc}
\usepackage[utf8x]{inputenc}
\usepackage{url}
\usepackage{xspace}
\usepackage{hyperref}
\usepackage{fancyvrb}
\input{macros} 
\usepackage{keyval}
\usepackage{fancyvrb}
\usepackage{color}
\usepackage{float}
\usepackage{ifthen}
\usepackage{calc}
\usepackage{ifplatform}
\usepackage{listings}
\input{pygmentize}

\usepackage{mymacros}

\def\mz{\lstinline}
\def\qmz|#1|{``\mz|#1|''}

\title{The ins and outs of iteration in \mezzo}

\authorinfo{Armaël Guéneau}
           {ENS Lyon \& INRIA}
           {armael.gueneau@inria.fr}
\authorinfo{François Pottier}
           {INRIA}
           {francois.pottier@inria.fr}
\authorinfo{Jonathan Protzenko}
           {INRIA}
           {jonathan.protzenko@ens-lyon.org}

\begin{document}
\maketitle

\begin{abstract}
This is a talk proposal for HOPE 2013. Using iteration over a collection as a
case study, we wish to illustrate the strengths and weaknesses of the
prototype programming language \mezzo.

\end{abstract}

\section{Introduction}

\mezzo~\cite{pottier-protzenko-13} is a high-level programming language in the
style of~ML. It is equipped with a strong static discipline of duplicable and
affine permissions, which controls aliasing and ownership, and rules out
certain mistakes, such as representation exposure and data races. In this
talk, we would like to illustrate how \mezzo expresses transfers of ownership:
sometimes easily, sometimes less so. We use iteration, a surprisingly rich problem,
as a case study.

\section{Algebraic data structures}

Thanks to algebraic data types, it is easy to define list- and tree-like
data structures. For instance, here is a type of mutable binary trees:
\begin{Verbatim}[commandchars=\\\{\}]
\PY{k}{data} \PY{k}{mutable} \PY{n}{tree} \PY{n}{a} \PY{o}{=}
  \PY{n+nc}{Leaf}
\PY{o}{|} \PY{n+nc}{Node} \PY{o}{\PYZob{}} \PY{n}{left}\PY{o}{:} \PY{n}{tree} \PY{n}{a}\PY{o}{;} \PY{n}{elem}\PY{o}{:} \PY{n}{a}\PY{o}{;} \PY{n}{right}\PY{o}{:} \PY{n}{tree} \PY{n}{a} \PY{o}{\PYZcb{}}
\end{Verbatim}

As in ML, a tree carries a tag, which identifies it as a leaf or a node.
Unlike in ML, a type is interpreted not just as a structural description, but
also as an assertion of ownership. Thus, when one writes \qmz|t @ tree a|,
this does not mean that ``\mz|t| is a tree (now and forever)''. Instead, this
means ``\mz|t| is a tree (now) and I have exclusive permission to read and
write it''. We say that \qmz|t @ tree a| is an affine permission: it is a
unique token that grants access to \mz|t| as a tree. A function that takes an
argument~\mz|t| and wishes to access it as a tree requests this token from its
caller and usually returns it to its caller. Permissions exist at
type-checking time and incur no runtime overhead.

The permission \qmz|t @ tree a| can be refined, by analysis of~\mz|t|,
into the permission ``\texttt{t @ Node \{ left: tree a; elem: a; right: tree a
\}}'', which itself
is automatically split by the type-checker into a conjunction of four
permissions:
\begin{Verbatim}[commandchars=\\\{\}]
\PY{n}{t} \PY{o}{@} \PY{n+nc}{Node} \PY{o}{\PYZob{}} \PY{n}{left} \PY{o}{=} \PY{n}{l}\PY{o}{;} \PY{n}{elem} \PY{o}{=} \PY{n}{x}\PY{o}{;} \PY{n}{right} \PY{o}{=} \PY{n}{r} \PY{o}{\PYZcb{}} \PY{o}{*}
\PY{n}{l} \PY{o}{@} \PY{n}{tree} \PY{n}{a} \PY{o}{*} \PY{n}{x} \PY{o}{@} \PY{n}{a} \PY{o}{*} \PY{n}{r} \PY{o}{@} \PY{n}{tree} \PY{n}{a}
\end{Verbatim}
(where \mz|l|, \mz|x|, \mz|r| are fresh auxiliary names).
Conjunction is naturally separating, so the
left and right subtrees must be disjoint: this really is
a type of trees, not of arbitrary graphs.

When reasoning abstractly, the permission \qmz|x @ a| is considered affine.
This allows the type variable~\mz|a| to be later instantiated with an affine
type, i.e., one that has a non-trivial interpretation in terms of
ownership. For instance, \qmz|t @ tree (ref int)| means that \mz|t| is a tree
of pairwise distinct integer references, and represents the ownership of the
tree and of its elements.

\section{Higher-order iteration}
\label{sec:hoiter}

The type-checker can split permissions (as above), join them, and set them
aside when they are not needed (a ``frame rule''). This makes it easy to write
a recursive function that descends into a tree. For instance, the type of the
``tree size'' function is:
\begin{Verbatim}[commandchars=\\\{\}]
\PY{k}{val} \PY{n}{size}\PY{o}{:} \PY{o}{[}\PY{n}{a}\PY{o}{]} \PY{n}{tree} \PY{n}{a} \PY{o}{\PYZhy{}}\PY{o}{\PYZgt{}} \PY{n}{int}
\end{Verbatim}
(Square brackets denote universal quantification.)
By convention, this means that the call \qmz|size t| requires the permission
\qmz|t @ tree a|
and returns it. It is equally easy to write a higher-order function that
descends into a tree and invokes a client-supplied function at every node:
\begin{Verbatim}[commandchars=\\\{\}]
\PY{k}{val} \PY{n}{iter}\PY{o}{:} \PY{o}{[}\PY{n}{a}\PY{o}{,} \PY{n}{s}\PY{o}{:} \PY{k}{perm}\PY{o}{]}
 \PY{o}{(}\PY{n}{f}\PY{o}{:} \PY{o}{(}    \PY{n}{a} \PY{o}{|} \PY{n}{s}\PY{o}{)} \PY{o}{\PYZhy{}}\PY{o}{\PYZgt{}} \PY{n}{bool}\PY{o}{,}
  \PY{n}{t}\PY{o}{:} \PY{n}{tree} \PY{n}{a} \PY{o}{|} \PY{n}{s}\PY{o}{)} \PY{o}{\PYZhy{}}\PY{o}{\PYZgt{}} \PY{n}{bool}
\end{Verbatim}
The function~\mz|f| has access to one tree element at a time: it receives a
permission of the form \qmz|x @ a| and must return it.  Thus, this element is
temporarily ``borrowed'' from the tree. The function~\mz|f| cannot access the
tree, since it does not receive a permission for it.
The universal quantification over a permission~\mz|s|, which~\mz|f| receives
and returns, and which~\mz|iter| also receives and returns, allows the client
to supply a function~\mz|f| that has a side effect on an area of memory
represented by~\mz|s|.
The Boolean value returned by~\mz|f| indicates whether iteration should
continue (i.e., \mz|false| represents an early termination request). The
Boolean value returned by~\mz|iter| indicates whether iteration went all
the way to the end (i.e., \mz|false| means iteration was terminated early).

\section{Tree iterators as an abstract data type}
\label{sec:adt}

The higher-order function \mz|iter| offers a style of iteration where the
provider invokes the consumer when an element is available. In contrast, some
programming languages, such as Java, encourage a style where the consumer
invokes the provider in order to obtain an element. In \mezzo, at present, it
is possible to encode this idiom, but this requires a deep understanding of
the system. Expressing the iterator \emph{interface} is tricky, because:
\begin{enumerate}
\item the permission for the collection (here, a tree) must disappear while
  the iterator is active, and must somehow be recovered once the iterator
  is discarded;
\item the permission for an element that was yielded by the iterator must be
  surrendered before the iterator can be queried again.
\end{enumerate}
Furthermore, writing an iterator \emph{implementation} is tricky, because:
\begin{enumerate}
\item[3.] whereas the function \mz|iter| relies on an implicit control stack,
  an iterator contains an explicit representation of this stack, whose shape
  must be described by an appropriate permission---typically, some kind of
  ``tree segment''.
\end{enumerate}

An iterator interface, in the form of an abstract data type (ADT) equipped
with a number of operations, can be expressed as follows. The type
\mz|tree_iterator| is parameterized with the type~\mz|a| of the elements and
with a permission~\mz|post|, which represents the underlying collection. The
idea is that this permission is recovered when the iterator is discarded.
\begin{Verbatim}[commandchars=\\\{\}]
\PY{k}{abstract} \PY{n}{tree\PYZus{}iterator} \PY{n}{a} \PY{o}{(}\PY{n}{post}\PY{o}{:} \PY{k}{perm}\PY{o}{)}
\end{Verbatim}

A tree iterator is created by invoking the function \mz|new|:
\begin{Verbatim}[commandchars=\\\{\}]
\PY{k}{val} \PY{n}{new}\PY{o}{:} \PY{o}{[}\PY{n}{a}\PY{o}{]}
  \PY{o}{(}\PY{k}{consumes} \PY{n}{t}\PY{o}{:} \PY{n}{tree} \PY{n}{a}\PY{o}{)} \PY{o}{\PYZhy{}}\PY{o}{\PYZgt{}}
  \PY{n}{tree\PYZus{}iterator} \PY{n}{a} \PY{o}{(}\PY{n}{t} \PY{o}{@} \PY{n}{tree} \PY{n}{a}\PY{o}{)}
\end{Verbatim}

When the type-checker examines the function call \qmz|let it = new t in ...|,
it checks that the permission \qmz|t @ tree a| is available at the program
point before the call. At the program point after the call, this permission is
gone (as specified by the \mz|consumes| keyword), and the permission \qmz|it @
tree_iterator a (t @ tree a)| appears instead.

At any moment, one can discard the iterator and recover the
permission~\mz|post| (which, in this context, is \qmz|t @ tree a|) by
invoking~\qmz|stop it|:
\begin{Verbatim}[commandchars=\\\{\}]
\PY{k}{val} \PY{n}{stop}\PY{o}{:} \PY{o}{[}\PY{n}{a}\PY{o}{,} \PY{n}{post}\PY{o}{:} \PY{k}{perm}\PY{o}{]}
  \PY{o}{(}\PY{k}{consumes} \PY{n}{it}\PY{o}{:} \PY{n}{tree\PYZus{}iterator} \PY{n}{a} \PY{n}{post}\PY{o}{)} \PY{o}{\PYZhy{}}\PY{o}{\PYZgt{}} \PY{o}{(}\PY{o}{|} \PY{n}{post}\PY{o}{)}
\end{Verbatim}
One may hope (and this holds in our implementation) that \mz|stop| has no
runtime effect and runs in constant time. It may be possible to extend \mezzo
with a notion of ``ghost'' code and to declare \mz|stop| as a ghost function.

While an iterator is active, it can be queried for a new element:
\begin{Verbatim}[commandchars=\\\{\}]
\PY{k}{val} \PY{n}{next}\PY{o}{:} \PY{o}{[}\PY{n}{a}\PY{o}{,} \PY{n}{post}\PY{o}{:} \PY{k}{perm}\PY{o}{]}
  \PY{o}{(}\PY{k}{consumes} \PY{n}{it}\PY{o}{:} \PY{n}{tree\PYZus{}iterator} \PY{n}{a} \PY{n}{post}\PY{o}{)} \PY{o}{\PYZhy{}}\PY{o}{\PYZgt{}}
  \PY{n}{either} \PY{o}{(}\PY{n}{focused} \PY{n}{a} \PY{o}{(}\PY{n}{it} \PY{o}{@} \PY{n}{tree\PYZus{}iterator} \PY{n}{a} \PY{n}{post}\PY{o}{)}\PY{o}{)}
         \PY{o}{(}\PY{o}{|} \PY{n}{post}\PY{o}{)}
\end{Verbatim}
The call \qmz|next it| requires \qmz|it @ tree_iterator a post|,
and (perhaps surprisingly) consumes this permission, which means that, immediately after
this call, the iterator can no longer be used. One must first examine the
value returned by the call. Roughly speaking, either:
\begin{enumerate}
\item it carries an element~\mz|x| of type~\mz|a|, together with a promise
  that by abandoning the permission \qmz|x @ a|, one can recover
  \qmz|it @ tree_iterator a post| and continue using the iterator; or
\item it carries the permission~\mz|post|, because the iterator has stopped.
\end{enumerate}
(We omit the definition of the algebraic data type \mz|either|, which 
represents a binary sum.)
A value of type~\qmz|focused a post| can be thought of as
a dependent pair of
a value~\mz|x| of
type~\mz|a| and a ``magic wand'', that is, a ``one-shot'' ability to
convert~\qmz|x @ a| to \mz|post|:
\begin{Verbatim}[commandchars=\\\{\}]
\PY{k}{alias} \PY{n}{focused} \PY{n}{a} \PY{o}{(}\PY{n}{post}\PY{o}{:} \PY{k}{perm}\PY{o}{)} \PY{o}{=}
  \PY{o}{(}\PY{n}{x}\PY{o}{:} \PY{n}{a}\PY{o}{,} \PY{n}{release}\PY{o}{:} \PY{n}{wand} \PY{o}{(}\PY{n}{x} \PY{o}{@} \PY{n}{a}\PY{o}{)} \PY{n}{post}\PY{o}{)}
\end{Verbatim}
\mezzo does not currently have a primitive concept of a magic wand, viewed as
a permission. As an approximation, we view a magic wand of \mz|pre|
to \mz|post| as a runtime function
that consumes~\mz|pre| and produces~\mz|post|. (Again, ideally, this
should be a ghost function.)
We make it a ``one-shot''
function (i.e., one that can be invoked at most once) by specifying
that it consumes an abstract permission \mz|ammo| and by pairing it with just
one copy of~\mz|ammo|. (Curly braces denote existential
quantification. An abstract permission is by default considered affine.)
\begin{Verbatim}[commandchars=\\\{\}]
\PY{k}{alias} \PY{n}{wand} \PY{o}{(}\PY{n}{pre}\PY{o}{:} \PY{k}{perm}\PY{o}{)} \PY{o}{(}\PY{n}{post}\PY{o}{:} \PY{k}{perm}\PY{o}{)} \PY{o}{=}
  \PY{o}{\PYZob{}}\PY{n}{ammo}\PY{o}{:} \PY{k}{perm}\PY{o}{\PYZcb{}} \PY{o}{(}
    \PY{o}{(}\PY{o}{|} \PY{k}{consumes} \PY{o}{(}\PY{n}{pre} \PY{o}{*} \PY{n}{ammo}\PY{o}{)}\PY{o}{)} \PY{o}{\PYZhy{}}\PY{o}{\PYZgt{}} \PY{o}{(}\PY{o}{|} \PY{n}{post}\PY{o}{)}
  \PY{o}{|} \PY{n}{ammo}\PY{o}{)}
\end{Verbatim}

This concludes the definition of the iterator interface.
We do not show the implementation of tree iterators, but note that the
internal definition of \mz|tree_iterator| itself relies on \mz|focused|:
\begin{Verbatim}[commandchars=\\\{\}]
\PY{k}{alias} \PY{n}{tree\PYZus{}iterator} \PY{n}{a} \PY{o}{(}\PY{n}{post}\PY{o}{:} \PY{k}{perm}\PY{o}{)} \PY{o}{=}
  \PY{n}{ref} \PY{o}{(}\PY{n}{focused} \PY{o}{(}\PY{n}{list} \PY{o}{(}\PY{n}{tree} \PY{n}{a}\PY{o}{)}\PY{o}{)} \PY{n}{post}\PY{o}{)}
\end{Verbatim}
This means that a tree iterator is a stack of sub-trees and that, by abandoning
the ownership of this stack, one recovers~\mz|post|, which represents the
ownership of the complete tree.

\section{Generic iterators as objects}
\label{sec:ooiter}

We have constructed an abstract data type of iterators for a
specific type of trees. The same approach can be applied to
other data structures. Unfortunately, every time one does so,
one obtains a new abstract data type of iterators.
Thus, one cannot write generic code that uses ``an iterator'' without knowing
how this iterator was constructed.

One way out of this problem is to adopt an object-oriented (OO) style and to
define an iterator to be an object equipped with \mz|next| and \mz|stop|
methods. The methods must have access to the iterator's internal state,
which we represent by an abstract permission~\mz|s|. Thus, an iterator is a
package of two functions that require~\mz|s| and of~\mz|s| itself:
\begin{Verbatim}[commandchars=\\\{\}]
\PY{k}{data} \PY{n}{iterator\PYZus{}s} \PY{o}{(}\PY{n}{s}\PY{o}{:} \PY{k}{perm}\PY{o}{)} \PY{n}{a} \PY{o}{(}\PY{n}{post}\PY{o}{:} \PY{k}{perm}\PY{o}{)} \PY{o}{=}
  \PY{n+nc}{Iterator} \PY{o}{\PYZob{}}
    \PY{n}{next}\PY{o}{:} \PY{o}{(}\PY{o}{|} \PY{k}{consumes} \PY{n}{s}\PY{o}{)} \PY{o}{\PYZhy{}}\PY{o}{\PYZgt{}} \PY{n}{either} \PY{o}{(}\PY{n}{focused} \PY{n}{a} \PY{n}{s}\PY{o}{)}
                                   \PY{o}{(}\PY{o}{|} \PY{n}{post}\PY{o}{)}\PY{o}{;}
    \PY{n}{stop}\PY{o}{:} \PY{o}{(}\PY{o}{|} \PY{k}{consumes} \PY{n}{s}\PY{o}{)} \PY{o}{\PYZhy{}}\PY{o}{\PYZgt{}} \PY{o}{(}\PY{o}{|} \PY{n}{post}\PY{o}{)}
  \PY{o}{|} \PY{n}{s} \PY{o}{\PYZcb{}}

\PY{k}{alias} \PY{n}{iterator} \PY{n}{a} \PY{o}{(}\PY{n}{post}\PY{o}{:} \PY{k}{perm}\PY{o}{)} \PY{o}{=}
  \PY{o}{\PYZob{}}\PY{n}{s}\PY{o}{:} \PY{k}{perm}\PY{o}{\PYZcb{}} \PY{n}{iterator\PYZus{}s} \PY{n}{s} \PY{n}{a} \PY{n}{post}
\end{Verbatim}
This is an encoding in the style of Pierce and Turner, 
with the added twists that \mz|s| is a permission, not a type (so the client never
obtains a pointer to the object's internal state) and \mz|s| is affine (so the
client cannot invoke \mz|stop| twice, for instance).

An ADT-style iterator can be converted a posteriori to OO-style. This is done
by the following function, which accepts an iterator~\mz|it| of an arbitrary
type~\mz|i|, provided this type is equipped with appropriate \mz|next| and
\mz|stop| operations.
\begin{Verbatim}[commandchars=\\\{\}]
\PY{k}{val} \PY{n}{wrap}\PY{o}{:} \PY{o}{[}\PY{n}{a}\PY{o}{,} \PY{n}{i}\PY{o}{,} \PY{n}{post}\PY{o}{:} \PY{k}{perm}\PY{o}{]} \PY{o}{(}
  \PY{k}{consumes} \PY{n}{it}\PY{o}{:} \PY{n}{i}\PY{o}{,}
  \PY{n}{next}\PY{o}{:} \PY{o}{(}\PY{k}{consumes} \PY{n}{it}\PY{o}{:} \PY{n}{i}\PY{o}{)} \PY{o}{\PYZhy{}}\PY{o}{\PYZgt{}}
    \PY{n}{either} \PY{o}{(}\PY{n}{focused} \PY{n}{a} \PY{o}{(}\PY{n}{it} \PY{o}{@} \PY{n}{i}\PY{o}{)}\PY{o}{)} \PY{o}{(}\PY{o}{|} \PY{n}{post}\PY{o}{)}\PY{o}{,}
  \PY{n}{stop}\PY{o}{:} \PY{o}{(}\PY{k}{consumes} \PY{n}{it}\PY{o}{:} \PY{n}{i}\PY{o}{)} \PY{o}{\PYZhy{}}\PY{o}{\PYZgt{}}     \PY{o}{(}\PY{o}{|} \PY{n}{post}\PY{o}{)}
\PY{o}{)} \PY{o}{\PYZhy{}}\PY{o}{\PYZgt{}} \PY{n}{iterator} \PY{n}{a} \PY{n}{post}
\end{Verbatim}
By combining \mz|wrap| and the ADT-style library of the previous section (\sref{sec:adt}),
one obtains:
\begin{Verbatim}[commandchars=\\\{\}]
\PY{k}{val} \PY{n}{new\PYZus{}tree\PYZus{}iterator}\PY{o}{:} \PY{o}{[}\PY{n}{a}\PY{o}{]}
  \PY{o}{(}\PY{k}{consumes} \PY{n}{t}\PY{o}{:} \PY{n}{tree} \PY{n}{a}\PY{o}{)} \PY{o}{\PYZhy{}}\PY{o}{\PYZgt{}} \PY{n}{iterator} \PY{n}{a} \PY{o}{(}\PY{n}{t} \PY{o}{@} \PY{n}{tree} \PY{n}{a}\PY{o}{)}
\end{Verbatim}
Conversely, one can convert from OO style to ADT style, in the following
sense: if desired, the type \mz|iterator| defined above can be equipped
with three operations \mz|new| (a constructor), \mz|next|, and \mz|stop|,
and made abstract.

An OO-style iterator is a stream (with mutable internal state), so it should
not be surprising that many of the standard operations on streams can be
defined on the type \mz|iterator|. For instance,
\mz|filter| creates a new iterator out of an existing one:
\begin{Verbatim}[commandchars=\\\{\}]
\PY{k}{val} \PY{n}{filter}\PY{o}{:} \PY{o}{[}\PY{n}{a}\PY{o}{,} \PY{n}{p}\PY{o}{:} \PY{k}{perm}\PY{o}{,} \PY{n}{post}\PY{o}{:} \PY{k}{perm}\PY{o}{]} \PY{o}{(}
  \PY{k}{consumes} \PY{n}{it}\PY{o}{:} \PY{n}{iterator} \PY{n}{a} \PY{n}{post}\PY{o}{,}
  \PY{n}{f}\PY{o}{:} \PY{o}{(}\PY{n}{a} \PY{o}{|} \PY{n}{p}\PY{o}{)} \PY{o}{\PYZhy{}}\PY{o}{\PYZgt{}} \PY{n}{bool}
\PY{o}{|} \PY{k}{consumes} \PY{n}{p}\PY{o}{)} \PY{o}{\PYZhy{}}\PY{o}{\PYZgt{}} \PY{n}{iterator} \PY{n}{a} \PY{o}{(}\PY{n}{p} \PY{o}{*} \PY{n}{post}\PY{o}{)}
\end{Verbatim}
A few points are worth noting:
\begin{enumerate}
\item The pre-existing iterator is consumed. It becomes owned by the new
  iterator, so to speak. Stopping the new iterator transparently stops
  the underlying iterator and yields~\mz|post|, which represents the
  ownership of the underlying collection(s).
\item The function~\mz|f| may have internal state, represented by the
  permission~\mz|p|, which \mz|f| requires and returns.
  This permission is consumed by the call to~\mz|filter|
  (i.e., the new iterator takes possession of~\mz|p|), and
  is recovered when the new iterator stops.
\item \mz|f| receives a permission to examine an element of type~\mz|a|, and
  must return this permission (there is no \mz|consumes| keyword).
\end{enumerate}
Other examples of operations that can be expressed include \mz|map|, \mz|zip|,
\mz|concat|, \mz|equal| (which in the case of trees solves the ``same fringe''
problem), etc.

\section{Turning a fold inside out}

We have presented two approaches to iteration. In one, the producer
is in control and invokes the consumer via a function call
(\sref{sec:hoiter}); in the other, this situation is reversed (\sref{sec:adt}
and \sref{sec:ooiter}). The former approach makes it easy to implement a
producer, while the latter approach facilitates life for the consumer,
especially when one wishes to draw data out of several
collections simultaneously.

In order to get the best of both approaches, one would like to be able to
automatically derive a first-order iterator in the style of \sref{sec:adt} and
\sref{sec:ooiter} out of a higher-order iteration function in the style of
\sref{sec:hoiter}. It is well-known that this can in principle be achieved by
using control operators. Unfortunately, for the moment at least, \mezzo does
not have first-class control. Yet, as a preliminary study, we verify that an
iterator can be derived out of a higher-order iteration function
written in continuation-passing style (CPS).

In order to allow early termination, we use a ``double-barreled continuation'' style,
and work with pairs of an abort continuation and a normal continuation. The type
\mz|continuations pre b1 b2|, defined below, represents such a pair. The abstract
permission \mz|ammo| is analogous to the one that was used in the definition of
\mz|wand| (\sref{sec:adt}) so as to ensure that a magic wand is applied
at most once. Here, because both continuations require the same \mz|ammo|, and
because only copy of \mz|ammo| is included, it ensures that at most one of the two continuations
can be invoked.
\begin{Verbatim}[commandchars=\\\{\}]
\PY{k}{alias} \PY{n}{continuations} \PY{o}{(}\PY{n}{pre} \PY{o}{:} \PY{k}{perm}\PY{o}{)} \PY{n}{b1} \PY{n}{b2} \PY{o}{=}
  \PY{o}{\PYZob{}} \PY{n}{ammo} \PY{o}{:} \PY{k}{perm} \PY{o}{\PYZcb{}} \PY{o}{(}
    \PY{n}{failure}\PY{o}{:} \PY{o}{(}\PY{o}{|} \PY{k}{consumes} \PY{o}{(}\PY{n}{ammo} \PY{o}{*} \PY{n}{pre}\PY{o}{)}\PY{o}{)} \PY{o}{\PYZhy{}}\PY{o}{\PYZgt{}} \PY{n}{b1}\PY{o}{,}
    \PY{n}{success}\PY{o}{:} \PY{o}{(}\PY{o}{|} \PY{k}{consumes} \PY{o}{(}\PY{n}{ammo} \PY{o}{*} \PY{n}{pre}\PY{o}{)}\PY{o}{)} \PY{o}{\PYZhy{}}\PY{o}{\PYZgt{}} \PY{n}{b2}
  \PY{o}{|} \PY{n}{ammo}
  \PY{o}{)}
\end{Verbatim}
The two continuations have the same domain: they expect zero runtime argument
and they consume a permission~\mz|pre|, which is a parameter of this definition.
They may have distinct answer types, \mz|b1| and \mz|b2|.

A CPS version of \mz|iter| (\sref{sec:hoiter}) has the following type:
\begin{Verbatim}[commandchars=\\\{\}]
\PY{k}{val} \PY{n}{cps\PYZus{}iter}\PY{o}{:} \PY{o}{[}\PY{n}{a}\PY{o}{,} \PY{n}{s} \PY{o}{:} \PY{k}{perm}\PY{o}{,} \PY{n}{b1}\PY{o}{,} \PY{n}{b2}\PY{o}{]} \PY{o}{(}
  \PY{k}{consumes} \PY{n}{t}\PY{o}{:} \PY{o}{(}\PY{n}{tree} \PY{n}{a} \PY{o}{|} \PY{n}{s}\PY{o}{)}\PY{o}{,}
  \PY{n}{f}\PY{o}{:} \PY{o}{(}
    \PY{k}{consumes} \PY{n}{x}\PY{o}{:} \PY{o}{(}\PY{n}{a} \PY{o}{|} \PY{n}{s}\PY{o}{)}\PY{o}{,}
    \PY{k}{consumes} \PY{n}{continuations} \PY{o}{(}\PY{n}{x} \PY{o}{@} \PY{o}{(}\PY{n}{a} \PY{o}{|} \PY{n}{s}\PY{o}{)}\PY{o}{)} \PY{n}{b1} \PY{n}{b2}
  \PY{o}{)} \PY{o}{\PYZhy{}}\PY{o}{\PYZgt{}} \PY{n}{b2}\PY{o}{,}
  \PY{k}{consumes} \PY{n}{continuations} \PY{o}{(}\PY{n}{t} \PY{o}{@} \PY{o}{(}\PY{n}{tree} \PY{n}{a} \PY{o}{|} \PY{n}{s}\PY{o}{)}\PY{o}{)} \PY{n}{b1} \PY{n}{b2}
\PY{o}{)} \PY{o}{\PYZhy{}}\PY{o}{\PYZgt{}} \PY{n}{b2}
\end{Verbatim}
The original \mz|iter| (\sref{sec:hoiter}) does not consume
``\mz+t @ (tree a | s)+'': it requires this permission and returns it to its
caller. Here, this idea remains valid in principle. However, \mz|cps_iter|
does not return the permission ``\mz+t @ (tree a | s)+'' to its caller:
instead, it transmits this permission to one of its continuations. By the same
token, the callback function~\mz|f| requires the permission ``\mz+x @ (a | s)+''
and transmits this permission to one of its continuations.

Using \mz|cps_iter|, one can re-implement \mz|new_tree_iterator| (\sref{sec:ooiter}).
The idea is to apply \mz|cps_iter| to a function \mz|yield| that
captures the current continuation pair and stores it within the iterator.
We omit the (fairly interesting) details. It is worth noting that this
construction is independent of trees, so, by abstracting over
\qmz|tree a| and \mz|cps_iter|, it can be turned into a re-usable library.

\section{Other approaches to iteration}

One particularly elegant approach, well-known in the functional programming
community, is to view the producer as a function that returns a lazy stream of
elements, while the consumer is a function that accepts such a stream. In this
approach, both producer and consumer are written in direct style, and the
transfer of control is implicit. In \mezzo, by building upon a primitive
notion of lock, one can define a type \qmz|thunk a| of suspensions, and on top
of that, a type \qmz|stream a| of lazy streams. Suspensions and streams are
considered duplicable, which means that they can be shared without
restriction, just as in ML or Haskell. This approach works well, but is
restricted to the case where the type~\mz|a| is itself duplicable.

Another attractive approach consists in running the producer and consumer as
two threads connected by channels. The communication protocol is as
follows. The producer sends an element~\mz|x| of type~\mz|a| along one
channel, and the consumer replies (once it is done processing this element) by
sending a message of type~\mz+(| x @ a)+ (i.e., no runtime value, and the
permission for~\mz|x|) along a second channel. Thus, the reply channel must be
heterogeneous: every message has a different type, as it concerns a different
element~\mz|x|. A limited kind of heterogeneous channel can be axiomatized in
\mezzo, but this topic deserves further study: taking inspiration from Villard
\etal.'s work~\citeyear{villard-lozes-calcagno-09}, we would like to understand
how to best express complex communication protocols in \mezzo and how to extend
\mezzo's formal proof of type soundness with these features.

\section{Conclusion}

In this talk proposal, we have refrained from including numerous citations.
However, there is a rich literature on type systems for mutable state and on
approaches to iteration. Krishnaswami \etal.'s
paper~\citeyear{krishnaswami-design-patterns-09} is particularly relevant.
We will draw a comparison with some of the related work during the talk.

The ``little'' problem of iteration is surprisingly rich, because it involves
the two fundamental issues of modularity and transfer of ownership.
We believe that it is a good way of illustrating \mezzo's expressive power
and, more generally, how one might program in a language equipped with a
pervasive notion of permission.

\sloppy
\hbadness=10000
\bibliographystyle{plain}
\bibliography{english}

\end{document}